\documentstyle[12pt]{article}
\begin{document}
\title{ POSSIBILITY OF GEOMETRIC DESCRIPTION
OF QUASIPARTICLES IN SOLIDS}
\author{
YU.A.~DANILOV, A.V.~ROZHKOV, and V.L.~SAFONOV$^{*}$}
\date{}
\maketitle

\centerline{\it Institute of Molecular Physics,
Russian Research Center "Kurchatov Institute",}

\centerline{\it 123182 Moscow, Russia}
\par

\medskip
\begin{abstract}
New  phenomenological  approach   for   the   description   of
elementary collective  excitations  is  proposed.  The  crystal  is
considered to be an anisotropic space-time vacuum with a prescribed
metric  tensor  in  which  the   information   on   electromagnetic
crystalline fields is included. The quasiparticles  in  this  space
are supposed to be described by the equations structurally  similar
to the relativistic wave equations for particles in empty space.
The generalized  Klein-Gordon-Fock  equation  and  the  generalized
Dirac equation in external electromagnetic  field  are  considered.
The  applicability  of  the  proposed  approach  to  the  case   of
conduction electron in a crystal is discussed.
\end{abstract}
\medskip
\noindent
PACS Nos.: 71.45 , 72 , 11.10.Q
\bigskip
\bigskip
\par \noindent $^{*}$ Corresponding author.
\par
E-mail: danilov@lmpp.kiae.su

\newpage
\section{ Motivation}
\par
The  concept  of  a  quasiparticle  as  a  quantum  of   collective
excitation  of  a  many-particle  system  is  of   extremely   high
importance  in  solid  state  physics.   In   many   respects   the
quasiparticle behavior resembles that of usual  particle  in  empty
space, but the peculiarity of  a  crystal  has  an  impact  on  the
dynamics of its elementary  excitations.  "The  properties  of  any
quasiparticles reveal the properties  of  their  areal  --  crystal,
i.e. atoms, molecules,  and  ions,  periodically  distributed  over
space" $^{1}$.
\par
Because  a  quasiparticle  in  its  essence  is  a  result  of
collective perturbation of all atoms the crystal consists  of,  the
crystal symmetry can not but affect the quasiparticle structure and
dynamics, and it seems  to  be  natural  to  study  all  elementary
collective excitations starting  with  the  most  general  symmetry
considerations.
\par
A model in which a crystal is considered to  be  a  continuous
anisotropic medium turns out to be enough simple  and  fruitful.  A
general approaches making possible  the  study  of  long-wavelength
collective excitations  in  the  frame  of  this  model  have  been
developed for elastic waves (and, consequently, for the  quanta  of
elastic excitations - phonons)$^{2}$  and  for  spin  waves  (and  their
quanta -- magnons)$^{3,4}$.
\par
Being restricted only to translational degrees of  freedom,  a
conduction electron (and a hole) in a crystal  with  inversion  can
be described in the long-wave approximation in the frame of a model
with effective dynamical mass tensor $m_{\alpha \beta }$ $\;^{4-6}$.
This tensor involves
the unknown internal forces exerting  on  a  quasiparticle  by  the
lattice. The wave function of such quasiparticle is  so  structured
that its small-scale details are smeared due to  the  averaging:  a
crystal reveals itself only as some anisotropic physical space with
metric tensor $\gamma _{\alpha \beta } = m_{\alpha \beta }/m$ ,
where $m$  is a mass of free electron.
\par
Below we shall mainly focus our attention to the case of  spin
1/2 quasiparticles. Traditional approach  takes  into  account  the
crystal anisotropy introducing a potential of periodic  crystalline
field in  non-relativistic  Hamiltonian  of  "bare"  electron  with
relativistic terms required to meet spin and orbital motions.  This
means that the  "dressing  up"  of  an  electron  begins  with  the
equation resulting from the Dirac equation with built-in  spherical
symmetry  of  empty   space,   perturbed   by   switching   on   of
electromagnetic field. It is easy  to  understand  that  with  such
approach  the  breaking  of  "being   on   equal   terms"   between
translational and spin motions is introduced {\it ab initio}.
In spite of the fact that for the most part of  known  effects
related with conduction electrons the  influence  of  spin  motions
is negligibly small, the  role  of  the  latters  can  considerably
enhance for strong anisotropic quasi-two-dimensional and
quasi-one-dimensional structures. So the problem is
what  to  be  understood
under spin of a particle (and, naturally, quasiparticle).

\medskip

\subsection{ What is Spin of a Quasiparticle?}
\par
From  one  hand  we  can  consider  spin  to   be   some   internal
rotational motion of a point particle. In this case it is simple to
imagine  a  picture  of  one-  and  two-dimensional  motion  of  an
electron-quasiparticle, such that its spin magnetic moment  equally
reacts to the magnetic field in any direction of  three-dimensional
space. Such a picture is enough popular  among  physicists  inspite
that the rotation of a dimensionless point hardly can be understood
from the viewpoint of common sense.
\par
On the other hand, one can keep to the point of view  proposed
by  Belinfante $^{7,8}$   according  to  which  spin  of  electron   is
completely similar to the angular momentum of classical  circularly
polarized wave. In this case the anisotropy  of  crystalline  space
has to be {\it ab initio} taken into account  in  the  equation  both  in
translational and in spin motion of quasiparticle (the  generalization
of Belinfante's theorem is given in Appendix). It  is  obvious
that in this case any rotation in one-dimensional motion is out  of
question, i.e. spin magnetic moment  of  electron-quasiparticle  in
one-dimensional case must be equal  zero,  and  in  two-dimensional
moment only component of spin magnetic moment perpendicular to  the
plane of motion remains finite.
\par
Here we share the second point of view.  We  can  mention  the
results of recent experimental works $^{9,10}$ which count in its favor.
They imply that in quasi-twodimensional crystalline  systems  (such
as quantum  well  and  layered  crystal)  only  component  of  spin
magnetic moment of quasiparticles,  transversal  to  the  plane  of
motion, reacts to the external magnetic field.

\medskip

\subsection{ Geometrical Approach to Description of Quasiparticles}
\par
So  we  suppose  that   the   influence   of   crystalline   medium
anisotropy must be taken into account  both  in  translational  and
spin motion of quasiparticles. In order to carry out such procedure
it is possible to make a natural assumption that the  quasiparticle
wave function in the above mentioned anisotropic space with  metric
tensor $\gamma _{\alpha \beta }({\bf x})$ (containing  the  unknown
information   on   the
crystalline field) satisfy the equation structurally similar to the
Dirac equation. In other words, with such approach {\it a  quasiparticle
is a particle in anisotropic space} -- anisotropic vacuum, which role
is fulfilled by crystal. The foundations of such approach were laid
in Refs.11-13, where  for  the  sake  of  simplicity  a  model  of
uniform anisotropic space was  considered  which  applicability  to
crystal  is  restricted  by  long-wavelength  states.   Preliminary
results for the case of nonuniform space were published in  Ref.14.
It  should  be  noted   that   the   phenomenological   geometrical
considerations were always attractive in physics (see, for example,
Wheeler's geometrodynamics $^{15}$ ).
\par
The plan of the work is as follows. First, keeping in the mind
methodical aims, we consider the generalized relativistic  equation
for quasiparticle with zero spin. Then we discuss the applicability
of the effective mass method  for  space  with  metric  tensor.  In
section 3 we study the generalized Dirac equation for quasiparticle
with spin 1/2. The analysis given in the work and its  consequences
are discussed in conclusion.

\medskip

\section{ Equation for Spin 0 Particle in Anisotropic Space}
\par
The Klein-Gordon-Fock equation for spin 0 particle in empty space
is

\begin{equation}
\left( \hbar \over c \right)^{2} {\partial^2 \over \partial t^2}\; \phi -
\hbar ^2 \Delta \phi = - m^{2}c^{2} \phi  .
\end{equation}
\smallskip
\par
The generalized Klein-Gordon-Fock equation for the space  with
metric tensor $g_{ij}(t,{\bf x})$  can be written as follows $^{16}$ :

\begin{equation}
\hbar ^{2}{1 \over \sqrt{-g}}\; {\partial \over \partial x_i}( g_{ij}
\sqrt{-g}\; {\partial \over \partial x_j } ) \phi  = - m^{2}c^{2} \phi.
\end{equation}
\smallskip
\par \noindent
Here $g \equiv  \det (g^{ij})$,  and the summation rule over  repeated
indices $(i,j = \overline{0,3} )$ is supposed. It is obvious that substituting
$g_{ij} =  \eta _{ij} \equiv$ diag $(1,-1,-1,-1)$ in (2), we obtain (1).
\par
Further we shall restrict ourselves with the representation of
metric tensor as

\begin{equation}
g_{ij} = \pmatrix{ 1 & 0 \cr 0 & {-\gamma_{\alpha \beta} }} ,
\qquad \alpha , \beta  = \overline{1,3},
\end{equation}
\par \noindent
where $\gamma _{\alpha \beta }({\bf x})$  is three-dimensional metric tensor,
determining  the
geometrical properties of the space. Then (2) can be  rewritten  as
follows

\begin{equation}
\left( {\hbar \over c} \right) ^{2} {\partial ^2 \over \partial t^2}\; \phi -
{\hbar ^2 \over \sqrt{\gamma}}\; {\partial \over \partial x_{\alpha}}
\left( \gamma _{\alpha \beta } \sqrt{\gamma}\; {\partial \over \partial x_{\beta }}
\right) \phi = - m^{2}c^{2} \phi ,
\end{equation}
\smallskip
\par \noindent
where $\gamma  \equiv  \det (\gamma ^{\alpha \beta }) = - g$ , and the summation
is carried  out  over repeated indices.
\par
The interaction between particle and external  electromagnetic
fields, described by potential $A^{i} = (\varphi , {\bf A})$,  is taken into
account by changing of the momentum operator $\hat{p}^{i}$  for generalized
momentum operator $\hat{P}^{i} \equiv  \hat{p}^{i} + (e/c) A^{i}$, where $e$
is the  charge  of  particle.
Then we obtain from (4)
\smallskip

\begin{eqnarray}
&\left( {\hbar \over c} \right) ^{2}
\left( {\partial \over \partial t} + ie \varphi \right)^2 \phi  -   \\
&{\hbar ^{2} \over \sqrt{\gamma}} \left(
{\partial \over \partial x_{\alpha }} - i {e \over \hbar c} A^{\alpha} \right)
\left( \gamma _{\alpha \beta } \sqrt{\gamma}\; (
{\partial \over \partial x_{\beta }} - i {e\over \hbar c} A^{\beta} ) \right)\phi
= - m^{2}c^{2} \phi. \nonumber
\end{eqnarray}

\smallskip

\subsection{ Non-relativistic Limit}
\par
\noindent Let us substitute in (5) $\phi  = \exp
( - imc^{2}t / \hbar ) \phi ^\prime $  and suppose  kinetic
and interaction energies of a particle to be  small  in  comparison
with the rest energy.  Disregarding  the  corresponding  terms,  we
obtain:
\begin{equation}
i \hbar  {\partial \phi ^\prime \over \partial t} =
{1 \over  2m \sqrt{\gamma}} ( \hat{P}^{\alpha } - {e\over c} A^{\alpha })
\left( \gamma _{\alpha \beta } \sqrt{\gamma } \; (
\hat{P}^{\beta } - {e\over c} A^{\beta }) \right) \phi ^\prime + e\varphi \; \phi ^\prime .
\end{equation}
\smallskip
\par \noindent
Let us emphasize that the derived equation has the  aspect  of  the
Schr\"odinger equation with Hamiltonian
\begin{equation}
{\cal H} = - {\hbar ^{2} \over 2m} \tilde{\Delta } + {e \over mc}
\gamma _{\alpha \beta } A^{\alpha } P^{\beta } - {ie\hbar \over mc}
(\tilde{\nabla }_{\beta } A^{\beta }) + {e^{2} \over 2mc^2} \gamma _{\alpha \beta}
A^{\alpha } A^{\beta } + e\varphi  ,
\end{equation}
\smallskip
\par \noindent
which is Hermitian operator in respect to scalar product $(\phi ,\psi ) =
\int  \phi ^{*}\psi \sqrt{\gamma }\; d^{3}x $ . Here we keep the following
notations
$$
\tilde{\Delta } \equiv  \tilde{\nabla }_{\beta } \tilde{\nabla }^{\beta }
= {1 \over \sqrt{\gamma }}\; {\partial \over \partial x_{\alpha}}
(\gamma _{\alpha \beta } \sqrt{\gamma }\; {\partial \over \partial x_{\beta }})
$$
and
$$
\tilde{\nabla }_{\beta } \equiv  {1 \over \sqrt{\gamma }}\;
{\partial \over \partial x_{\alpha }} ( \gamma _{\alpha \beta }
\sqrt{\gamma }).
$$

\smallskip

\subsection{ Translation-invariant Metrics}
\par
Consider the Hamiltonian (7) in the  case  when  the  space  metric
tensor is invariant in  respect  to  some  full  crystalline group:
\begin{equation}
\gamma _{\alpha \beta }({\bf x}+{\bf l}) = \gamma _{\alpha \beta }({\bf x}) ,
\end{equation}
\begin{equation}
{\cal A}^{\mu }_{\alpha } {\cal A}^{\nu }_{\beta } \gamma _{\mu \nu }
(\hat{{\cal A}}^{-1}{\bf x}) = \gamma _{\alpha \beta }({\bf x}) ,
\end{equation}
\par \noindent
where ${\bf l}$ is a vector from the translation group of the lattice, and
$\hat{{\cal A}} = \Vert {\cal A}^{\alpha }_{\beta } \Vert $  is transformation
matrix belonging to the  point  group of the lattice.
\par
For zero external field one has

\begin{equation}
\hat{{\cal H}}_{0}= - {\hbar ^{2} \over 2m} {1 \over \sqrt{\gamma }}
{\partial \over \partial x_{\alpha }} ( \gamma _{\alpha \beta } \sqrt{\gamma}\;
{\partial \over \partial x_{\beta }} ).
\end{equation}
\par \noindent
Due to properties of tensor $\gamma _{\alpha \beta }({\bf x})$ ,  this
Hamiltonian  has  the
symmetry of crystalline lattice: it  commutes  with  all  operators
from full  crystalline  symmetry  group.  Because  the  full  group
contains the group of discrete  translations, the eigenfunctions of
such Hamiltonian can be classified  according  to  the  irreducible
representations of this group. Let $\hat{T}_{{\bf l}}$  is operator
of translation
by  the  lattice  vector ${\bf l}$ , $\Psi ({\bf x})$   is  eigenfunction
of  the Hamiltonian (10) then
$$
\hat{T}_{{\bf l}}f({\bf x}) = f({\bf x}+{\bf l}), \qquad
\hat{{\cal H}}_{0}\Psi ({\bf x}) = {\cal E}\Psi ({\bf x}),
$$
and
$$
[\hat{T}_{{\bf l}},\hat{{\cal H}}_{0}] \equiv
\hat{T}_{{\bf l}}\hat{{\cal H}}_{0} - \hat{{\cal H}}_{0}\hat{T}_{{\bf l}} = 0.
$$
\smallskip
\par \noindent
It follows from the last equation that
$$
0=(\hat{T}_{{\bf l}}\hat{{\cal H}}_{0} - \hat{{\cal H}}_{0}\hat{T}_{{\bf l}})
\Psi ({\bf x})=\hat{T}_{{\bf l}}\hat{{\cal H}}_{0}\Psi ({\bf x})-
\hat{{\cal H}}_{0}\hat{T}_{{\bf l}}\Psi ({\bf x})
$$
i.e.
$$
\hat{{\cal H}}_{0}(\hat{T}_{{\bf l}}\Psi ({\bf x})) = {\cal E}
\hat{T}_{{\bf l}}\Psi ({\bf x}).
$$
\smallskip
\par
It can be seen from here that $(\hat{T}_{{\bf l}}\Psi ({\bf x}))$ in  the
same  way  as
$\Psi ({\bf x})$ is eigenfunction of $\hat{{\cal H}}_{0}$ , and the space
of the  eigenfunctions
reffering to a given eigenvalue ${\cal E}$, can be expanded into the direct
sum of the spaces-supports of irreducible  representations  of  the
group  of  discrete  translations.  Due  to  the  fact   that   the
translation  group  of  the  lattice  is   commutative,   all   its
representations are one-dimensional, i.e. the eigenfunctions can be
chosen in such way that for any $\Psi ({\bf x})$:
$\hat{{\cal H}}_{0}\Psi ({\bf x}) = {\cal E}\Psi ({\bf x})$  there  exist
complex numbers $\xi _{{\bf l}}$,  $\mid \xi _{{\bf l}}\mid  = 1$ ,  such
that $\hat{T}_{{\bf l}}\Psi ({\bf x}) = \xi _{{\bf l}}\Psi ({\bf x})$ .  For
$\xi _{{\bf l}}$  the relation $\xi _{{\bf l}}\xi _{{\bf l^\prime }} =
\xi _{{\bf l+l^\prime }}$  is valid which  follows  from  the
chain of equations
\smallskip
$$
\hat{T}_{{\bf l}}\hat{T}_{{\bf l^\prime }}\Psi ({\bf x}) =
\hat{T}_{{\bf l+l^\prime }}\Psi ({\bf x}) = \xi _{{\bf l}}\xi _{{\bf l^\prime }}
\Psi ({\bf x}) = \xi _{{\bf l+l^\prime }}\Psi ({\bf x}).
$$
\smallskip
\par \noindent
Therefore, $\xi _{{\bf l}} = \exp (-i{\bf kl})$,  where ${\bf k}$  is a
vector  determining  the
irreducible representation of the translation group,  according  to
which $\Psi ({\bf x})$  transforms: $\Psi ({\bf x}) \equiv
\Psi _{{\bf k}}({\bf x})$ ,  $\hat{T}_{{\bf l}}\Psi _{{\bf k}}({\bf x}) =
\exp (-i{\bf kl})\Psi _{{\bf k}}({\bf x})$.
It is easy to see that $\Psi _{{\bf k}}({\bf x})$  can be written as

\begin{equation}
\Psi _{{\bf k}}({\bf x}) = \exp (-i{\bf kx})u_{{\bf k}}({\bf x}),
\end{equation}
\noindent where $u_{{\bf k}}({\bf x})$   is  periodic  function  with  the
lattice  period, normalized to the condition

\begin{equation}
\int \mid u_{{\bf k}}\mid ^{2}\sqrt{\gamma}\; d^{3}x = 1 ,
\end{equation}
\noindent where the integration is carried out over the volume of  elementary
cell.
\par
The relations (11) and (12) extends the Bloch theorem  to  the
case of the spaces with non-euclidean metrics.

\medskip

\subsection{ Effective Mass Method}
\par
Let potential $\varphi ({\bf x})$  slightly vary over the distances  of  the
order of elementary cell.  Consider  the  matrix  elements  of  the
Hamiltonian $\hat{{\cal H}} = \hat{{\cal H}}_{0}+ e\varphi $  in the Bloch
wave representation
$$
{\cal H}_{{\bf kk^\prime }}= \int \Psi ^{*}_{{\bf k^\prime }}(\hat{{\cal H}}_{0}
+ e\varphi )\Psi _{{\bf k}}\sqrt{\gamma}\; d^{3}x = {\cal E}_{{\bf k}}
\delta _{{\bf kk^\prime }}+ \int \; e^{-i({\bf k}-{\bf k^\prime }){\bf x}}
e\varphi  (u^{*}_{{\bf k^\prime }}u_{{\bf k}}\sqrt{\gamma})\; d^{3}x.
$$
\noindent Because $\mid {\bf k} - {\bf k^\prime \mid }$  must be small in
comparison with the size of the reciprocal lattice cell,
the rapidly oscillating  factor $(u^{*}_{{\bf k^\prime }}u_{{\bf k}}
\sqrt{\gamma})$ can be changed for its average

\begin{equation}
{\cal H}_{{\bf kk^\prime }}= {\cal E}_{{\bf k}}\delta _{{\bf kk^\prime }}+
\int u^{*}_{{\bf k^\prime }}u_{{\bf k}}\sqrt{\gamma}\; d^{3}x \;
\int \; e^{-i({\bf k}-{\bf k^\prime }){\bf x}} e\varphi \; d^{3}x .
\end{equation}
\smallskip
\noindent Let us consider the integral
$$
\int  u^{*}_{{\bf k^\prime }}u_{{\bf k}}\sqrt{\gamma}\; d^{3}x = 1 +
({\bf k^\prime }- {\bf k}) \int (\nabla _{{\bf k}}u^{*}_{{\bf k}}) u_{{\bf k}}
\sqrt{\gamma}\; d^{3}x + \ldots \quad ,
$$
\smallskip
\noindent where $\nabla _{{\bf k}}$  means differentiation by ${\bf k}$ .
In  this  expansion  by
powers of $\;{\bf k^\prime }- {\bf k}\;$  the terms of the higher than the first
order, and due to (12) the value of integral for
$\;{\bf k^\prime }= {\bf k}\;$  is equal to 1.
\par
Let us show that if the Hamiltonian is invariant in respect to
inversion, then
$$
\int  u_{{\bf k}}\nabla _{{\bf k}}u^{*}_{{\bf k}}
\sqrt{\gamma}\; d^{3}x =  0 \qquad  {\rm and}  \qquad
\int  u^{*}_{{\bf k^\prime }}u_{{\bf k}}\sqrt{\gamma}\; d^{3}x =  1
$$
up to the terms of the second order in respect to ${\bf k^\prime }- {\bf k}$ .
Because the Hamiltonian
$\hat{{\cal H}}_{0}$  is real-valued, the following relations are valid
$$
\hat{{\cal H}}_{0}\Psi ^{*}_{{\bf k}}({\bf x}) = {\cal E}_{{\bf k}}
\Psi ^{*}_{{\bf k}}({\bf x}), \quad  \hat{{\cal H}}_{0}\Psi ^{*}_{{\bf k}}
(-{\bf x}) = {\cal E}_{{\bf k}}\Psi ^{*}_{{\bf k}}(-{\bf x}) .
$$
\smallskip
\noindent Let us determine now the wave quasimomentum
$\Psi ^{*}_{{\bf k}}(-{\bf x})$ . To  this  aim
let us act on it with the shift operator $\hat{T}_{{\bf l}}$ :
$$
\hat{T}_{{\bf l}}\Psi ^{*}_{{\bf k}}(-{\bf x}) = \Psi ^{*}_{{\bf k}}
(-{\bf x}-{\bf l}) = \left( \exp (i{\bf kl})\Psi _{{\bf k}}({\bf x})\right) ^{*}=
\exp (-{i{\bf kl}})\Psi ^{*}_{{\bf k}}({\bf x}).
$$
\smallskip
\noindent Whence $\Psi ^{*}_{{\bf k}}(-{\bf x}) = \Psi _{{\bf k}}({\bf x})$
and, therefore,

\begin{equation}
u^{*}_{{\bf k}}(-{\bf x}) = u_{{\bf k}}({\bf x}).
\end{equation}
\noindent Then
$$
u^{*}_{{\bf k}}(-{\bf x}) \nabla _{{\bf k}}u_{{\bf k}}(-{\bf x}) =
u_{{\bf k}}({\bf x}) \nabla _{{\bf k}}u^{*}_{{\bf k}}({\bf x}) ,
$$
\smallskip
$$
\int  u^{*}_{{\bf k}}(-{\bf x}) \nabla _{{\bf k}}u_{{\bf k}}(-{\bf x})
\sqrt{\gamma}\; d^{3}x = \int  u_{{\bf k}}({\bf x}) \nabla _{{\bf k}}
u^{*}_{{\bf k}}({\bf x})\sqrt{\gamma}\; d^{3}x,
$$
\smallskip
$$
\int  u^{*}_{{\bf k}}({\bf x}) \nabla _{{\bf k}}u_{{\bf k}}({\bf x})
\sqrt{\gamma }\; d^{3}x = \int  u_{{\bf k}}({\bf x}) \nabla _{{\bf k}}
u^{*}_{{\bf k}}({\bf x})\sqrt{\gamma }\; d^{3}x
$$
$$
= {1\over 2} \int \left( {u_{{\bf k}}({\bf x}) \nabla _{{\bf k}}
u^{*}_{{\bf k}}({\bf x}) + u^{*}_{{\bf k}}({\bf x})\nabla _{{\bf k}}u_{{\bf k}}
({\bf x})}\right) \sqrt{\gamma }\; d^{3}x
$$
$$
= {1\over 2} \nabla _{{\bf k}}\left( \int u_{{\bf k}}({\bf x})u^{*}_{{\bf k}}
({\bf x})\sqrt{\gamma }\;d^{3}x \right) = 0 .
$$
\smallskip
\noindent Thus

\begin{equation}
{\cal H}_{{\bf kk^\prime }} = {\cal E}_{{\bf k}}\delta _{{\bf kk^\prime }}+
\int \; e^{-i({\bf k}-{\bf k^\prime }){\bf x}} e\varphi \; d^{3}x
\end{equation}
\smallskip
\noindent up to the terms of the order $({\bf k}-{\bf k^\prime })^{2}$.
It can be  seen  from  (15)
that matrix elements of the perturbed Hamiltonian $\hat{{\cal H}}$
depend  only
on the Fourier-transform of the perturbing potential  and  spectrum
of non-perturbed Hamiltonian. Therefore two unperturbed  Hamiltonians
with the same or close spectra ${\cal E}_{{\bf k}}$   have  in  slowly
changing electric field the same eigenvalues and eigenfunctions in the Bloch
wave representation. The same idea underlies the  classical  method
of effective masses: the unperturbed Hamiltonian $\hat{{\cal H}}_{0}$
with spectrum
$$
{\cal E}_{k} = {\hbar ^2 \over 2} {\partial ^{2}{\cal E}_{k}\over \partial
k_{\alpha }\partial k_{\beta }} k_{\alpha }k_{\beta } + \ldots ,
$$
\noindent is changed for the Hamiltonian with effective masses
$$
\hat{{\cal H}}^\prime  = {1\over 2} (m^{-1}_{\hbox{eff}})^{\alpha \beta }
\hat{p}_{\alpha }\hat{p}_{\beta } , \quad (m^{-1}_{\hbox{eff}})^{\alpha \beta }
\equiv  {1\over \hbar ^{2}} {\partial ^{2}{\cal E}_{k}\over
\partial k_{\alpha }\partial k_{\beta }}.
$$

\medskip

\section{ Equation for Spin 1/2 Particle in Anisotropic Space}
\par
The  search  for  the  equation  to  be  found  must  be  obviously
started from the well-known Dirac equation
\smallskip

\begin{equation}
\left( i\hbar \gamma _{i} {\partial \over \partial x_i} -
mc \right) \psi  = 0 ,
\end{equation}
\smallskip
\noindent describing  particle  with  spin  1/2  in  spherically  symmetrical
space-time ($\gamma _{i}$ are Dirac's matrices).
\par
It is possible to assume  that  in  order  to  generalize  the
equation to the case of the  motion  of  a  particle  in  a  curved
space-time it is sufficient to change  partial  derivatives
$\partial /\partial x_{i}$
for contravariant ones $D^{i}$.  However, here the  difficutlies  ensue:
the spinor quantities $\psi $   are  not  tensor  objects,  which  makes
impossible the definition for them "good"  operation  of  covariant
differentiation.
\par
The procedure of covariant  differentiation  for  spinors  was
defined by Fock and Iwanenko $^{17}$ .  These authors  have  written  the
generalized Dirac equation for the case of the particle  motion  in
classical gravitational field. In this section the method developed
in Ref.17 is employed for the description of the particle motion in
the field of forces of non-gravitational origin.  Recall  that  the
action of electromagnetic forces is introduced  via  metric  tensor
depending on point of space.
\par
The  main  requirement  on  the  construction   of   covariant
derivative of a spinor is that  the  observable  tensor  quantities
bilinear in respect to $\psi $  could be  translated  as  tensors  in  a
curved space-time.
\par
Let  us  introduce  in  each  point  of  a   space-time   four
orthonormalized vectors -- tetrade $h^{(a)}_{i}(t,{\bf x})$,
$a = \overline{0,3}$ , $i = \overline{0,3}$ ,
and conjugate tetrade $h_{(a)i}(t,{\bf x})$ :
$$
h^{(a)}_{i}h_{(a)j} = \delta _{ij}\cdot diag(1,-1,-1,-1).
$$
The summation over repeated tetrade indices is supposed:
$$
f^{(a)}g_{(a)} = \sum^{3}_{a=0}f^{(a)}g_{(a)}.
$$
Shifting  tetrade  vectors  from  point
$x_{i}$   to
infinitesimally close point $x_{i}+ \delta s_{(a)}h^{(a)}_{i}$,  we  obtain
a  set  of
quantities $\gamma ^{bca}$ ,  $\gamma ^{bca}= -\gamma ^{cba}$ ,
called Ricci rotation coefficients.
The antisymmetric matrices $\gamma ^{{bca}} \delta s_{(a)}$ are matrices
of infinitesimal
rotation  bringing  into   coincidence   the   tetrade   in   point
$x_{i}+ \delta s_{(a)}h^{(a)}_{i}$ with  the  tetrade  shifted  from $x_{i}$  .
The  Ricci
coefficients can be expressed in the terms of covariant derivatives
of tetrade vectors $D_{k}h^{(a)}_{i} \equiv  h^{(a)}_{i;k}$ :
$$
\gamma ^{{bca}} = h^{(b)}_{i;k} h^{(c)i} h^{(a)k}.
$$
Thus the operator of infinitesimal parallel shift of  a  spinor $\psi $
can be expressed in terms of $\gamma ^{{bca}}$ :

\begin{equation}
D^{i}\psi  = \left( {\partial \over \partial x_i}+ {1\over 4} \gamma
^{{bca}}\gamma _{b}\gamma _{c}h^{i}_{(a)}\right) \psi  .
\end{equation}
\noindent As a result, the  general-covariant  generalization  of  the  Dirac
equation can be written as follows $^{16,17}$ :

\begin{equation}
\left( i \hbar \gamma _{a}h^{(a)}_{k}D^{k} - mc \right) \psi  = 0 .
\end{equation}
\noindent This equation depends explicitly  on  the  tetrade $h^{(a)}_{k}$
and  is
invariant only in respect to  local  Lorentzian  transformation  of
tetrade vectors.
\par
Because further we shall restrict ourselves to the case  where
the metric tensor is of type (3), let us introduce tetrade  vectors
in such way that
$$
h^{(0)}_{k}= (1,0,0,0),\quad h^{(a)}_{k}= (0, h^{(a)}_{1},h^{(a)}_{2},h^{(a)}_{3}),
\quad a = \overline{1,3} .
$$
\noindent For such choice of vectors, Ricci coefficient $\gamma ^{{bca}}$
vanishes if at
least one of indices $a$, $b$  or $c$  is equal to zero. Therefore, the
equation (18) can be written as

\begin{equation}
\left( {i\hbar \over c}\;\gamma _{0}\;{\partial \over \partial t}+i\hbar \gamma _{a}
h_{(a)\beta }D^{\beta }-mc \right) \psi  = 0 .
\end{equation}
\noindent Here and further the summation over repeated indices  is  supposed:
$$
\gamma _{a}h_{(a)} \equiv \sum^{3}_{a=1} \gamma _{a}h_{(a)}.
$$
The lowering of index is carried out according
to the following rule: $f_{(a)}= - f^{(a)}$.   This  rule  implies  that
$h^{\alpha }_{(a)}h^{\beta }_{(a)}= \gamma ^{\alpha \beta }$  and
$h^{\beta }_{(a)}h_{(b)\beta } = \delta _{ab}$ . The  equation  (19)  can  be
written as

\begin{equation}
i\hbar {\partial \psi \over \partial t} = \left(
i\hbar c\alpha _{a}h_{(a)\nu }D^{\nu }+mc^{2}\beta \right) \psi .
\end{equation}
where $\alpha _{a} \equiv  \gamma _{0}\gamma _{a}$ , $\beta  \equiv  \gamma _{0}$.
\par
Coming back to the operators $D^{\nu }$ , we obtain

\begin{eqnarray}
D^{\nu } = {\partial \over \partial x_{\nu }}-{1\over 4} \gamma _{{bca}}
\gamma _{b}\gamma _{c}h^{\nu }_{(a)} = {\partial \over \partial x_{\nu }} -
{1\over 4} \gamma _{{bca}}h^{\nu }_{(a)}\pmatrix{\sigma_b \sigma_a & 0 \cr
0 & \sigma_b \sigma_c } & \nonumber \\
= {\partial \over \partial x_{\nu }} - {i\over 4} \epsilon _{{bcd}}
\gamma _{{bca}}h^{\nu }_{(a)}\pmatrix{\sigma_d & 0 \cr 0 & \sigma_d } . &
\end{eqnarray}
\noindent Here $\sigma _{d}$  are Pauli matrices, $\epsilon _{{bcd}}$   is
completely  antisymmetric
tensor. In (21) the relation $\sigma _{a}\sigma _{b} = \delta _{ab}+
i\epsilon _{{abc}}\sigma _{c}$ , antisymmetricity
of Ricci coefficients $\gamma ^{{bca}} = - \gamma ^{{cba}}$ and
the rule of index lowering have been used.
\par
Introducing two two-component functions $\phi $  and $\chi $ :
$\psi  \equiv  \pmatrix{\phi \cr\chi }$ ,
one can bring the equation (20) to the form

\begin{eqnarray}
i\hbar {\partial \phi \over \partial t} = - i\hbar c\sigma _{a}h_{(a)\nu }
\hat{\nabla }^{\nu }\chi  + mc^{2}\phi , \nonumber \\
\; \\
i\hbar {\partial \chi \over \partial t} = - i\hbar c\sigma _{a}h_{(a)\nu }
\hat{\nabla }^{\nu }\phi  - mc^{2}\chi , \nonumber
\end{eqnarray}
\smallskip
where $\hat{\nabla }^{\nu } \equiv  {\partial \over \partial x_{\nu }} -
{i\over 4} \epsilon _{{bcd}}\gamma _{{bca}}h^{\nu }_{(a)}\sigma _{d}$ .
\par
\noindent Denoting
$$
\sigma _{a}h_{(a)\nu } \hat{\nabla }^{\nu } \equiv  ({\bf \sigma} ,\hat{{\bf \nabla}}),
$$
\noindent one obtains from (22)

\begin{eqnarray}
i\hbar {\partial \phi \over \partial t} = - i\hbar c({\bf \sigma} ,\hat{{\bf \nabla }})\chi
+ mc^{2}\phi , \nonumber \\
\; \\
i\hbar {\partial \chi \over \partial t} = - i\hbar c({\bf \sigma} ,\hat{{\bf \nabla }})\phi
- mc^{2}\chi . \nonumber
\end{eqnarray}
\smallskip
\par
The interaction with external electromagnetic field can be taken
into account by changing the derivatives ${\partial / \partial x_i}$
for ${\partial / \partial x_i} + (i e / \hbar c) A^{i}$.
Then the equation (23) is transformed to

\begin{eqnarray}
i\hbar {\partial \phi \over \partial t} = - i\hbar c({\bf \sigma} ,\hat{{\bf \nabla }}
+ i {e\over \hbar c} {\bf A})\chi  + e\varphi \cdot \phi  + mc^{2}\phi , \nonumber \\
\; \\
i\hbar {\partial \chi \over \partial t} = - i\hbar c({\bf \sigma} ,\hat{{\bf \nabla }}
+ i {e\over \hbar c} {\bf A})\phi  + e\varphi \cdot \chi - mc^{2}\chi . \nonumber
\end{eqnarray}

\medskip

\subsection{ Non-relativistic Limit}
\par
Let us find now the analogue of the Pauli equation assuming that
the rest energy $mc^{2}$  of  the  particle  considerably  exceeds  its
kinetic energy and the energy of interaction. To this  aim  let  us
make in (24) the following substitutions:
$$
\phi  = \exp \left( - {mc^{2}\over \hbar }t \right) \phi ^\prime  \quad
{\rm  and } \quad \chi  = \exp \left( - {mc^{2}\over \hbar }t \right)\chi ^\prime .
$$
\noindent Then

\smallskip
\begin{eqnarray}
i\hbar {\partial \phi^\prime \over \partial t} = - i\hbar c({\bf \sigma} ,
\hat{{\bf \nabla }} + i {e\over \hbar c} {\bf A})\chi ^\prime + e\varphi \cdot \phi ^\prime , \\
\; \nonumber \\
i\hbar {\partial \chi^\prime \over \partial t} = - i\hbar c({\bf \sigma} ,
\hat{{\bf \nabla }} + i {e\over \hbar c} {\bf A})\phi ^\prime + e\varphi \cdot \chi ^\prime - 2mc^{2}\chi ^\prime .
\end{eqnarray}
\smallskip
\noindent It follows from (26) that
$$
\chi ^\prime \simeq  - {i\hbar \over 2mc} ({\bf \sigma} ,\hat{{\bf \nabla}} +
i {e\over \hbar c} {\bf A})\phi ^\prime .
$$
\par
\noindent Substituting the latter expression in (25), we obtain

\begin{equation}
i\hbar {\partial \phi^\prime \over \partial t} = - {\hbar ^{2}\over 2m}
({\bf \sigma} ,\hat{{\bf \nabla }} + i {e\over \hbar c} {\bf A})^{2}\phi ^\prime + e\varphi \cdot \phi ^\prime .
\end{equation}
\smallskip
\noindent This equation has the aspect  of  the  Schr\"odinger  equation  with
Hamiltonian

\begin{equation}
\hat{{\cal H}} = - {\hbar ^{2}\over 2m} ({\bf \sigma} ,\hat{{\bf \nabla }}
+ i {e\over \hbar c} {\bf A})^{2}\phi ^\prime + e\varphi \cdot \phi ^\prime .
\end{equation}
\noindent Operator $\hat{{\cal H}}$  is Hermitian in respect to the scalar
product
$$
(\phi ,\psi ) = \int  \phi ^{+} \psi \sqrt{\gamma } \; d^{3}x ,
$$
\noindent where $\phi ^{+}$  is the conjugate spinor.
\par
Let us put the expression $({\bf \sigma} ,\hat{{\bf \nabla }} +
i {e\over \hbar c} {\bf A})^{2}$ in more detailed form:

\begin{eqnarray}
&({\bf \sigma} ,\hat{{\bf \nabla }} + i {e\over \hbar c} {\bf A})^{2}\phi  =
\sigma _{e}h_{(e)\nu }{\cal D}^{\nu }\sigma _{f}h_{(f)\mu }{\cal D}^{\mu }\phi  \\
&-{i\over 4} [ \sigma _{e}h_{(e)\nu }\epsilon _{bcd}\gamma _{bca}
h^{\nu }_{(a)}\sigma _{d}\sigma _{f}h_{(f)\mu }{\cal D}^{\mu }  \nonumber \\
& + \sigma _{e}
h_{(e)\nu }{\cal D}^{\nu }\sigma _{f}h_{(f)\mu }\epsilon _{ghi}
\gamma _{ghj}h^{\mu }_{(j)}\sigma _{i} ] \phi  \nonumber \\
& - {1\over 16} \sigma _{e}h_{(e)\nu }\epsilon _{bcd}\gamma _{bca}
h^{\nu }_{(a)}\sigma _{d}\sigma _{f}h_{(f)\mu }\epsilon _{ghi}
\gamma _{ghj}h^{\mu }_{(j)}\sigma _{i} \phi  ,  \nonumber
\end{eqnarray}
\smallskip
\noindent where
${\cal D}^{\beta } \equiv {\partial / \partial x_{\delta }} +
(i e / \hbar c) A^{\delta } $ .
\par
\noindent For the first summand in (29) it is valid that

\begin{eqnarray}
&\sigma _{e}h_{(e)\nu }{\cal D}^{\nu }\sigma _{f}h_{(f)\mu }{\cal D}^{\mu }\phi  =  & \\
&\sigma _{e}\sigma _{f}h_{(e)\nu }h_{(f)\mu }\left( D^{\nu }+
i{e\over \hbar c}A^{\nu } \right) {\cal D}^{\mu }\phi  + \sigma _{e}\sigma _{f}
h_{(e)\nu }h^{\quad \nu }_{(f)\mu ;} {\cal D}^{\mu }\phi  = & \nonumber \\
&\left( D^{\nu }+i{e\over \hbar c}A^{\nu } \right){\cal D}_{\nu }\phi +
i\epsilon _{{efc}}\sigma _{c}h_{(e)\nu }h_{(f)\mu }\left( D^{\nu }+
i{e\over \hbar c}A^{\nu }\right) {\cal D}^{\mu }\phi  & \nonumber \\
&+ \sigma _{e}\sigma _{f}h_{(e)\nu }h^{\quad \nu }_{(f)\zeta ;}
h^{\zeta }_{(a)}h_{(a)\mu }{\cal D}^{\mu }\phi  = & \nonumber \\
&{1\over \sqrt{\gamma }} \left( {\partial \over \partial x_{\nu }} +
i {e\over \hbar c} A^{\nu }\right) \gamma _{\nu \mu } \sqrt{\gamma} \; \left(
{\partial \over \partial x_{\mu }} + i {e\over \hbar c} A^{\mu }\right) \phi - & \nonumber \\
& {e\over \hbar c} H_{\nu }h^{\nu }_{(c)}\sigma _{c}\phi
 - \gamma _{{aff}}h_{(a)\mu }{\cal D}^{\mu }\phi  +
i\epsilon _{{fec}}\sigma _{c}\gamma _{{afe}}h_{(a)\mu }{\cal D}^{\mu }\phi  , & \nonumber
\end{eqnarray}
\smallskip
\par
\noindent where $H_{\beta } = ({1\over \sqrt{\gamma }} rot {\bf A})_{\beta }$
is component of external magnetic field.
\par
\noindent For the second summand it is valid that
\smallskip

\begin{eqnarray}
& - {i\over 4} \left( \sigma _{a}\sigma _{d}\sigma _{f}\epsilon _{{bcd}}
\gamma _{{bca}}h_{(f)\mu }{\cal D}^{\mu } + \sigma _{e}\sigma _{f}
\sigma _{i}\epsilon _{{ghi}}h_{(e)\mu }{\cal D}^{\mu }\gamma _{{ghf}}\right) \phi = &  \nonumber \\
& - {i\over 4} [ \sigma _{a}\sigma _{d}\sigma _{f}\epsilon _{{bcd}}
\gamma _{bca}h_{(f)\mu }{\cal D}^{\mu }+\sigma _{f}\sigma _{a}\sigma _{d}
\epsilon _{bcd}\gamma _{bca}h_{(f)\mu }{\cal D}^{\mu }  & \nonumber \\
& - \sigma _{e}\sigma _{f}\sigma _{i}\epsilon _{ghi}\gamma _{{ghf},e} ] \phi = & \\
& -{i\over 4} (\sigma _{a}\sigma _{d}\sigma _{f} + \sigma _{f}\sigma _{a}\sigma _{d})
\left( \epsilon _{{bcd}}\gamma _{{bca}}h_{(f)\mu }{\cal D}^{\mu }\right)\phi & \nonumber \\
& - {i\over 4}(\sigma _{e}\delta _{fi}+ i\epsilon _{{fic}}\sigma _{e}\sigma _{c})
\epsilon _{{ghi}}\gamma _{{ghf},e}\phi = & \nonumber \\
& -{i\over 4} ( 2\delta _{{ad}}\sigma _{f} + 2i\epsilon _{{adf}})
\left( \epsilon _{{bcd}}\gamma _{{bca}}h_{(f)\mu }{\cal D}^{\mu }\right) \phi \nonumber \\
& -{i\over 4} (\sigma _{e}\delta _{fi}- \epsilon _{{fic}}\epsilon _{{ecg}}
\sigma _{g} + i\epsilon _{{fie}}) \epsilon _{{ghi}}\gamma _{{ghf},e}\phi  = & \nonumber \\
& - {i\over 2} \epsilon _{{bca}}\gamma _{{bca}}h_{(f)\mu } {\cal D}^{\mu }\phi
+ {1\over 2}(\delta _{fb}\delta _{ac}-\delta _{fc}\delta _{ab})
\gamma _{{bca}}h_{(f)\mu }{\cal D}^{\mu }\; \phi  & \nonumber \\
& -{i\over 4} (\sigma _{e}\delta _{fi}- \sigma _{f}\delta _{ie} + \sigma _{i}\delta _{fe} +
i\epsilon _{{fie}}) \epsilon _{{ghi}}\gamma _{{ghf},e}\phi  = & \nonumber \\
& - {i\over 2} \epsilon _{{bca}}\gamma _{{bca}}h_{(f)\mu } {\cal D}^{\mu }\phi
+ \gamma _{{baa}}h_{(b)\mu } {\cal D}^{\mu }\phi  -
{i\over 4} \sigma _{e}(\epsilon _{{ghf}}\gamma _{{ghf},e} & \nonumber \\
& - \epsilon _{{ghf}}\gamma _{{ghe},f} + \epsilon _{{ghe}}\gamma _{{ghf},f})\phi
+ {1\over 2} \gamma _{{eff},e}\phi  . & \nonumber
\end{eqnarray}
\smallskip
\noindent And, at last, for the third summand in (29) it is valid that
\smallskip

\begin{eqnarray}
& - {1\over 16} \sigma _{a}\sigma _{d}\sigma _{f}\sigma _{i}(\epsilon _{{bcd}}
\gamma _{{bca}})(\epsilon _{{ghi}}\gamma _{{ghf}})\; \phi  = &  \\
& - {1\over 16} ( \delta _{{ad}}+ i\epsilon _{{adc}}\sigma _{c})( \delta _{fi}
+ i\epsilon _{{fij}}\sigma _{j})(\epsilon _{{bcd}}\gamma _{{bca}})
(\epsilon _{{ghi}}\gamma _{{ghf}})\; \phi  = & \nonumber \\
& - {1\over 16} (\delta _{{ad}}\delta _{fi} + i\delta _{fi}\epsilon _{{adc}}
\sigma _{c} + i\delta _{{ad}}\epsilon _{{fij}}\sigma _{j} & \nonumber \\
& - \delta _{af}\delta _{{di}}+ \delta _{fd}\delta _{ai})
(\epsilon _{{bcd}}\gamma _{{bca}})(\epsilon _{{ghi}}\gamma _{{ghf}})\phi  = & \nonumber \\
& - {1\over 16} [ (\epsilon _{{bca}}\gamma _{{bca}})^{2} + 2i(\epsilon _{{bca}}
\gamma _{{bca}})\epsilon _{{fij}}\epsilon _{{ghi}}\gamma _{{ghf}}\sigma _{j} & \nonumber \\
& - \epsilon _{{bcd}}\epsilon _{{ghd}}\gamma _{{bca}}\gamma _{{gha}} +
\epsilon _{{bcd}}\gamma _{{bca}}\epsilon _{{gha}}\gamma _{{ghd}} ] \phi  = & \nonumber \\
& - {1\over 16} [ (\epsilon _{{bca}}\gamma _{{bca}})^{2}+ 2i(\epsilon _{{bca}}
\gamma _{{bca}})(\delta _{gj}\delta _{fh}- \delta _{hj}\delta _{fg})\gamma _{{ghf}}
\sigma _{j} - &  \nonumber \\
& (\delta _{bg}\delta _{ch}- \delta _{bh}\delta _{cg})\gamma _{{bca}} \gamma _{{gha}}
+ \epsilon _{{bcd}} \gamma _{{bca}}\epsilon _{{gha}}\gamma _{{ghd}} ] \phi  =
[ -{1\over 16}(\epsilon _{{bca}}\gamma _{{bca}})^{2} & \nonumber \\
& - {i\over 4}(\epsilon _{{bca}}\gamma _{{bca}})\gamma _{{gff}}\sigma _{g}+
{1\over 8}\gamma _{{bca}}\gamma _{{bca}}- {1\over 16}\epsilon _{{bcd}}
\gamma _{{bca}}\epsilon _{{gha}}\gamma _{{ghd}} ] \phi . & \nonumber
\end{eqnarray}
\smallskip
\par
Collecting all terms (30)--(32), we obtain Hamiltonian (28)  in
its explicit form:
\smallskip

\begin{eqnarray}
& \hat{{\cal H}} = - {\hbar ^{2}\over 2m} {1\over \sqrt{\gamma }}\left(
{\partial \over \partial x_{\nu }} + i {e\over \hbar c} A^{\nu }\right)
\gamma _{\nu \mu }\sqrt{\gamma }\left( {\partial \over \partial x_{\mu }}
+ i {e\over \hbar c} A^{\mu }\right) & \\
& + i {\hbar ^{2} \over 4m} \epsilon _{{bca}}\gamma _{{bca}}h_{(f)\nu }
\sigma _{f}{\cal D}^{\nu } + {e\hbar \over 2mc} H_{\nu }h^{\nu }_{(c)}\sigma _{c} -
i {\hbar ^{2}\over 2m} \epsilon _{{fec}}\sigma _{c}\gamma _{{afe}}h_{(a)\mu }{\cal D}^{\mu } & \nonumber \\
& + {\hbar ^{2}\over 2m} \left( {1\over 16}(\epsilon _{{bca}}\gamma _{{bca}})^{2}-
{1\over 8} \gamma _{{bca}}\gamma _{{bca}}+
{1\over 16} \epsilon _{{bcd}}\gamma _{{bca}}\epsilon _{{gha}}\gamma _{{ghd}}-
{1\over 2} \gamma _{{eff},e}\right) & \nonumber \\
& + {i\hbar^2 \over 8m}\sigma _{c}(\epsilon _{{ghf}}\gamma _{{ghf},c}-
\epsilon _{{ghf}}\gamma _{{ghc},f} + \epsilon _{{ghc}}\gamma _{{ghf},f}+
\epsilon _{{bda}}\gamma _{{bda}}\gamma _{{cff}}) .& \nonumber
\end{eqnarray}

\medskip

\subsection{ Long-wavelength Approximation}
\par
The Hamiltonian (33) can be considerably simplified if  tensor $\gamma _{\alpha \beta }$
does not depend on the point of space. Such space is flat, and  the
metric tensor in this case can be brought to the diagonal form. The
description of quasiparticles  in  solid  state,  corresponding  in
non-relativistic limit to the effective mass method, for the  first
time was proposed in Refs.12,13. In this case the Hamiltonian  (33)
is as follows
\smallskip

\begin{equation}
\hat{{\cal H}} =\sum^{3}_{i=1} {1\over 2m_{i}} ( \hat{P}_{i}- {e\over c} A_{i} )^{2}
+ e\varphi  + {e\hbar \over 2c} \left( {H_{1}\sigma _{1}\over \sqrt{m_2 m_3}} +
{H_{2}\sigma _{2}\over \sqrt{m_3 m_1}} + {H_{3}\sigma _{3}\over \sqrt{m_1 m_2}}\right).
\end{equation}
\smallskip
\noindent One can see that the frequencies of spin  cyclotron  resonance  are
expressed in terms of  the  components $m_{i}$  of  effective  masses.
Similar relation was derived in the frame of standard  approach  by
Cohen and Blount$^{18}$ in the limit of strong spin-orbital interaction.
\par
Assuming that one of the effective mass turns  into  infinity,
the motion becomes  effectively  two-dimensional.  Only  the  field
component perpendicular to the plane of motion contributes  to  the
energy of interaction between spin and magnetic field. It should be
once more mentioned that this theoretical result  is  in  agreement
with the magnetotransport experiment of Martin {\it et al} $^{9}$  on strained
quantum  wells  and  with  the  normal   state   magnetothermopower
measurements of Jang {\it et al} $^{10}$  on high temperature  superconducting
$Nd_{1.85}Ce_{0.15}CuO_{4}$ crystal.
\par
If two masses turn into infinity, say $m_{1}$ and $m_{2}$ , then  only
the one-dimensional motion along $x_{3}$  axis survives, and the  terms
determining Zeeman interaction disappear.  In  the  frame  of  such
approximation one can not tell the electrons  with  opposite  spins
from each other, i.e. spin becomes a  hidden  coordinate.  On  this
base  in  Refs.12,13 a  hypothesis  has  been  put  forward   that
"one-dimensional" conduction electrons are governed  by  para-Fermi
statistics of the rank $p=2$ (the maximal occupation number). One can
expect that this para-Fermi statistics is also a good approximation
for quasi-one-dimensional case. This can follow from the continuity
of  physical   parameters   for   a   system   of   identical   and
quasi-identical particles.
\par
Concept of quasiparticles with parastatistics and  two  models
of superconducting systems with para-Fermi statistics (on the  base
of BCS model) were considered in Refs.19--21. The main conclusion is
that the critical temperature increases in system  with  para-Fermi
statistics. In Ref.22 the Fr\"ohlich's one-dimensional superconductor
with para-Fermi statistics has been considered.

\medskip

\section{ Conclusion}
\par
The given above analysis shows  the  self-consistency  of  the
proposed geometrical approach and traditionally  accepted  language
of solid state physics. The main aim of this work was to extend the
analogy of "particle-quasiparticle" to the  relativistic  case.  In
principle this way allows to consider spin properties of elementary
collective excitations. It is clear that such analysis  enables  to
consider the question on quantum statistics of quasiparticles  from
a new viewpoint.
\par
The question on spin (and spin relation to statistics) of such
quasiparticles, for  example,  phonons,  usually  does  not  arise,
because elementary excitations of  oscillatory  medium  unavoidably
are bosons. Essentially more complicated situation  we  meet  under
description of quasiparticles with half-integer spin. The matter is
the extension of Pauli's postulate on  Fermi  statistics  for  such
quasiparticles is by no means so obvious as  it  may  seem  to  be.
According to general theorem by Govorkov $^{23}$ on  connection  between
spin of particle and  statistics,  the  realization  of  para-Fermi
statistics for conduction electrons and holes can not be excluded {\it a
priori.}
\par
To  our  opinion  the  construction  of  unified   theory   of
elementary quasiparticles in  the  frame  of  only  electromagnetic
interaction  is  quite  possible  (the  problem   of   the   "Grand
Unification" does not arise).
\par
If the density of charged quasiparticles is not low, then  the
effects characteristic for quantum liquids come into play.  In  the
frame of proposed approach it is necessary to take into account the
change of metrics and as a consequence the change  of  the  problem
symmetry.
\par
Another important feature of the proposed approach is that  it
enables to understand the interconnection between  the  excitations
of "two-dimensional world" (of anyons and other planar  formations)
and real three-dimensional space.
\par
The  said  above  does  not  by  any  means  exhaust  the  all
possibilities of the proposed approach. For example, it  allows  to
describe the excitation in  fractal  media,  where  due  to  strong
anisotropy the description of  quasiparticles  drastically  changed
(instead of conventional differential equations the equations  with
fractional derivatives arise and instead of  smooth  solutions  one
meets fast oscillating functions and  distributions)  and  to  hope
that the question on the classification of  elementary  excitations
in fractal media can be settled as well.

\newpage

{\bf APPENDIX}

{\bf Generalized Belinfante's Theorem}
\par
Symmetric energy-momentum tensor for a spinor  field  has  the
form
$$
{\cal T}_{\mu \nu } = {i\hbar \over 4} ( \overline{\psi } \gamma _{\mu }\nabla _{\nu }\psi  + \overline{\psi } \gamma _{\nu }\nabla _{\mu }\psi  ) + \hbox{ h.c.} \qquad (A1)
$$
\noindent Corresponding density of momentum is
\begin{eqnarray*}
\pi _{k} = {\cal T}_{0k} = {i\hbar \over 4} ( \psi ^{+} \nabla _{k}\psi  +
{1\over c} \psi ^{+} \alpha _{k} {\partial \psi \over \partial t} ) + \hbox{ h.c. } \\
= {i\hbar \over 4} (\psi ^{+}\nabla _{k}\psi ) + {i\hbar \over 4} \psi ^{+}\alpha _{k}(\alpha ^{l}\nabla _{l} + mc\beta )\psi  +\hbox{ h.c. } \qquad (A2) \\
= {i\hbar \over 4} (\psi ^{+}\nabla _{k}\psi ) + {i\hbar \over 4} (\psi ^{+}\alpha _{k}\alpha ^{l}\nabla _{l}\psi ) +\hbox{ h.c.}
\end{eqnarray*}
\par \noindent
>From $ \alpha _{k}\alpha ^{l} + \alpha ^{l}\alpha _{k} = 2\delta ^{l}_{k} $  follows
$$
{i\hbar \over 4} (\psi ^{+}\alpha _{k}\alpha ^{l}\nabla _{l}\psi )
= {i\hbar \over 4} (\psi ^{+}\nabla _{k}\psi ) + {\hbar \over 4} \psi ^{+}\sigma ^{l}_{k}(\nabla _{l}\psi ) ,\qquad (A3)
$$
\noindent where $\sigma ^{l}_{k} = {i\over 2} [\alpha _{k},\alpha ^{l}]$.
Thus the density of momentum (A2) can be expressed as
$$
\pi _{k} = {i\hbar \over 2} [ \psi ^{+}(\nabla _{k}\psi ) - (\nabla _{k}\psi ^{+}) \psi  ] + {\hbar \over 4} \nabla _{l}(\psi ^{+}\sigma ^{l}_{k}\psi ) .\qquad (A4)
$$
\noindent Now we write the density of  moment  of  momentum  (as  it  was
proposed by Belinfante$^{7}$ for isotropic case) in the form
$$
j_{m} = {1\over \sqrt{\gamma }} \epsilon _{{mnk}} x^{n} \pi ^{k} = l_{m} + s_{m} ,\qquad (A5)
$$
$$
l_{m} = {1\over \sqrt{\gamma }} {i\hbar \over 2} \epsilon _{{mnk}} x^{n}
[ \psi ^{+}(\nabla _{l}\psi ) - (\nabla _{l}\psi ^{+}) \psi  ] \gamma ^{lk} ,
$$
$$
s_{m} = {1\over \sqrt{\gamma }}{\hbar \over 4} \epsilon _{{mnk}} x^{n} \nabla _{l}(\psi ^{+} \sigma ^{lk} \psi ) .
$$
\noindent Here $\gamma ^{lk} =$ diag $(m/m_{1}, m/m_{2}, m/m_{3})$, $\gamma  = \det (\gamma ^{lk})$.
One can see  that
$l_{m}$ does not depend on the spinor state which  is  typical  for  the
orbital moment and $s_{m}$ are defined by the  spinor  state.  The  last
term in (A5) can be rewritten as
$$
s_{m} = - {1\over \sqrt{\gamma }} {\hbar \over 2} (\psi ^{+} \sigma _{m} \psi ) = {i\over \sqrt{\gamma }} {\hbar \over 4} \epsilon _{{mkl}} \alpha ^{k}\alpha ^{l} ,\qquad (A6)
$$
\noindent where
$$
\sigma _{m} = {1\over 2} \epsilon _{{mlk}} \sigma ^{lk} = {i\over 2} \epsilon _{{mlk}} \alpha ^{k}\alpha ^{l} .
$$

\newpage

{\bf References}
\par \noindent
1. M.I.Kaganov and I.M.Lifshits,  {\it Quasiparticles}
(Nauka, Moscow, 1989, in Russian).
\par \noindent
2. J.A.Reissland, {\it The physics of phonons}  (Willey, London, 1973).
\par \noindent
3. A.F.Andreev and V.I.Marchenko,  {\it Usp. Fiz. Nauk} {\bf 130}, 39 (1980).
\par \noindent
4. E.M.Lifshits and L.P.Pitaevskii, {\it Statistical Physics}
(Nauka, Moscow, 1978, in Russian).
\par \noindent
5. H.J.Zeiger and G.W.Pratt,  {\it Magnetic interactions in solids}
(Clarendon Press, Oxford, 1973).
\par \noindent
6. N.B.Brandt and S.M.Chudinov,  {\it Electrons and phonons in metals}
(Moscow State University, Moscow, 1990, in Russian).
\par \noindent
7. F.G.Belinfante,  {\it Physica} {\bf 6}, 887 (1939).
\par \noindent
8. H.C.Ohanian,  {\it Amer. J. Phys.} {\bf 54}, 500 (1986).
\par \noindent
9. R.W.Martin, R.J.Nicholas, G.J.Rees, S.K.Haywood, N.J.Mason,
and P.J.Walker,  {\it Phys. Rev.} {\bf B42}, 9237 (1990).
\par \noindent
10. Wu Jiang, X.Q.Xu, S.J.Hagen, J.L.Peng, Z.Y.Li, and R.L.Greene,
{\it Phys. Rev.} {\bf B48}, 657 (1993).
\par \noindent
11. V.L.Safonov, {\it Phys. Stat. Sol. (b)} {\bf 171}, K19 (1992).
\par \noindent
12. V.L.Safonov, {\it Phys. Stat. Sol. (b)} {\bf 176}, K55 (1993).
\par \noindent
13. V.L.Safonov, {\it Int. J. Mod. Phys.} {\bf B7}, 3899 (1993).
\par \noindent
14. Yu.A.Danilov, A.V.Rozhkov, and V.L.Safonov,
{\it Preprint} IAE-5776/1 (RRC "Kurchatov Institute", Moscow, 1994).
\par \noindent
15. J.A.Wheeler,  {\it Neutrinos, gravitation and geometry}
(Rendiconti della Scuola Internazionale di Fisica "Enrico
Fermi"  Corso XI  pp.67-196. Bologna, 1960).
\par \noindent
16. A.A.Grib, S.G.Mamaev, and V.M.Mostepanenko,
{\it Vacuum quantum effects in strong fields}
(Energoatomizdat, Moscow, 1988, in Russian).
\par \noindent
17. V.Fock and D.Iwanenko,
{\it Comp. Rend. Acad. Sci. Paris} {\bf 188}, 1470 (1929).
\par \noindent
18. M.H.Cohen and E.I.Blount,  {\it Phil. Mag.} {\bf 5}, 115 (1960).
\par \noindent
19. V.L.Safonov, {\it Phys. Stat. Sol. (b)} {\bf 167}, 109 (1991).
\par \noindent
20. V.L.Safonov,  {\it Superconductivity (KIAE)} {\bf 3}, S380 (1990).
\par \noindent
21. V.L.Safonov, {\it Phys. Stat. Sol. (b)} {\bf 174}, 223 (1992).
\par \noindent
22. V.L.Safonov and A.V.Rozhkov, {\it Mod. Phys. Lett.} {\bf B8}, 1195 (1994).
\par \noindent
23.  A.B.Govorkov, {\it Teor. Mat. Fiz.} {\bf 98}, 163 (1994).
\end{document}